\documentclass[paper]{JHEP}
\usepackage{amssymb} 
\def\a{\begin{eqnarray}}
\def\b{\end{eqnarray}} 
\def\0{\nonumber} 
 
\preprint{SISSA/74/98/EP/FM\\hep-th/yymmxxx}
\title{Enhanced Gauge Symmetries on Elliptic $K3$}

\author{L.Bonora$^\dagger$, C.Reina, A.Zampa\\ 
International School for Advanced
Studies(SISSA/ISAS)\\ 
Via Beirut 2--4, 34014 Trieste, Italy\\
$^\dagger$ and INFN, Sezione di Trieste}

\abstract{We show that the geometry of K3 surfaces with singularities of 
type A-D-E contains enough information to reconstruct  a copy of the Lie 
algebra associated to the given Dynkin diagram.
We apply this construction to explain the enhancement of symmetry in F
and IIA theories compactified on singular K3's.}
 
\received{} 
\accepted{}
\begin{document}

\section{Introduction}

F--theory and IIA superstring theory compactified on a K3 are conjectured 
to be $s$--dual to heterotic string theory on a 2--torus and 4--torus, 
respectively.
At generic points of the moduli space of the latter, the gauge symmetry is
abelian while at special moduli the gauge group is non--abelian.  
F--theory and IIA theory do not possess non--abelian gauge fields in their
perturbative spectrum.
Therefore, if duality is to hold, something exceptional must happen
corresponding to special moduli in such a way that a non--abelian gauge symmetry
appears. This gauge symmetry enhancement is conjectured on the basis of duality
and is mostly supported by the appearance of Dynkin diagrams in correspondence
with the resolution of singular K3's.  However we would like to reverse this
argument and ask ourself whether the K3 geometry, in a IIA or F--theory
environment, contains enough information to allow us to retrieve a
non--abelian gauge theory framework. 
Accordingly, symmetry enhancement is not just a consequence of the 
conjectured duality, but rather constitutes a piece of evidence of it.

In this paper we address exactly this problem.  That is, we consider 
F--theory (or type IIA theory) compactified on an elliptic K3. We go to 
the limit in which the K3 becomes singular (where a non--abelian gauge 
symmetry is expected to arise) and we ask ourselves whether from these 
data we can reconstruct the framework of a non--abelian gauge theory.
We shall see that, in the case of a singular elliptic K3, it is in fact
possible to associate to singular fibres, whose singularity is of
A-D-E type, a copy of the Lie algebra with the same Dynkin diagram
\footnote{In this paper we limit ourselves to the simplest possible 
examples of symmetry enhancement. We do not consider here, for example, 
the appearance of non--simply laced symmetry groups or of `frozen' 
singularities, see \cite{witten1} and references therein.}.

\section{From duality to symmetry enhancement}

In this section we recall a few standard facts about heterotic string 
compactified on tori and the symmetry enhancements that ensue in 
F--theory or IIA theory compactified on K3 as a consequence of
the hypothesis of duality.

\subsection{Gauge symmetry of the heterotic string compactified on tori}

Let us consider for definiteness \cite{narain} the heterotic SO(32) 
superstring compactified on a torus $T^p$. The 
coordinates of the string are divided in three groups: the uncompactified 
coordinates will be denoted $X^\lambda$, the $X^I$ with $I=1,...,16$ are
left--moving scalars on the maximal torus of SO(32), while $X^i$ with 
$i=1,...,p$ are the compactified string coordinates.  $n^I,m^I$ and $n^i, 
m_i$ will be the corresponding winding and KK numbers. 
The general constant background (moduli) involves a gauge field $A_i^I$, 
beside the metric $g_{ij}$ and the two--form potential $B_{ij}$.
The lattice of the conjugate momenta, $\Gamma_{p,16+p}$, defined by 
\a 
&&\tilde L^i = n^i -B^i{}_k n^k+{1\over 2} g^{ij}m_j +{1\over 2}
g^{ij}A^I_j (m^I - {1 \over 2} A_k^In^k)\0\\ &&L^i=n^i +B^i{}_k
n^k-{1\over 2} g^{ij}m_j -{1\over 2} g^{ij}A^I_j (m^I - {1 \over 2}
A_k^In^k)\0\\ &&L^I = m^I - A^I_i n^i\label{lattice} 
\b 
has scalar product 
\a
L^i g_{ij}L^j-\tilde L ^i g_{ij} \tilde L'^j - L^IL'^I   = - n_im'^i - n'_i m^i
- m^I m'^I\label{quadratic} 
\b 
and is unimodular, integral and even.

The spectrum of the compactified theory is determined by the physicality
conditions in the different sectors of the theory.  The moduli space of the
theory contains $p^2$ parameters corresponding to $g_{ij}$ and $B_{ij}$, and
$16p$ parameters $A_i^I$.  For generic values of the parameters we have $16 +
2p$ massless vector states.  They are obtained by simply choosing $n^i= m_j=0$
and $m^I=0$ for any $i,j$ and $I$ (i.e.  we sit at the origin of the lattice)
and forming the right
\a
b_{-{1\over 2}}^\lambda|0>_R\otimes \tilde \alpha_{-1}^i|0>_L,\quad
b_{-{1\over 2}}^\lambda|0>_R\otimes \tilde
\alpha_{-1}^I|0>_L\}\label{cartan} 
\b 
and left states
\a 
b_{-{1\over 2}}^i|0>_R\otimes
\tilde \alpha_{-1}^\lambda|0>_L\label{cartan'} 
\b 
respectively.
The $\alpha$ oscillators are the bosonic ones, the $b(d)$ are the NS(R) 
fermionic ones, as usual. Therefore, at a generic point of the
moduli space one finds an abelian gauge group $U(1)^{16+2p}$.  More massless
states, and therefore possible enhancing of symmetries, can be found at
particular points of the moduli space. We give an explicit example in 
the Appendix.
There we show that at the point of the moduli space determined by 
$g_{11}=g_{22}= 1/4, g_{12}=0$ and $A_i^I= \delta_i^I$, the symmetry of 
the heteroric theory compactified on $T^2$ is enhanced to 
SO(36)$\times$U(1)$^2$. Choosing different backgrounds we can 
find an enormous variety of different gauge groups (of total rank 20).

One could have started from the ${\rm E_8\times E_8}$ heterotic string instead,
but once compactified on a torus the two heterotic theories are equivalent 
\cite{ginsparg}. 
Similar things can be repeated for the heterotic string compactified on $T^p$.
In this case the total rank of the gauge group is $16+2p$. In general the
moduli space of the theory is, apart from the dilaton, isomorphic to
\a
{\cal M}_{{\bf h},p} = O(p,16+p,{\mathbb{Z}})\setminus O(p,16+p, {\mathbb{R}})/
O(p,{\mathbb{R}})\times O(16+p, {\mathbb{R}})\0
\b
where $O(p,16+p, {\mathbb{Z}})$ represents the group of t--duality equivalences
\cite{GPR}.

\subsection{Gauge symmetry on $K3$}

In this subsection we summarize how one can figure gauge symmetry enhancement 
on the IIA and F--theory compactified on a K3 surface, on the ground that these
theories must be dual to the heterotic string theory on $T^4$ and $T^2$, 
respectively.

The basic observation behind duality is to identify part of the moduli 
space of the heterotic theory with the moduli space of suitable 
structures on the K3 surface. For example, in 
the case of IIA compactified on K3, the moduli space of Einstein metrics on K3
is embedded in the moduli space of the conformal non--linear $\sigma$--model on K3, which in
turn can be identified with ${\cal M}_{{\bf h},4}$, \cite{aspin}.
The moduli space of the Einstein--K\"ahler metrics on K3 surface $X$ is 
isomorphic to the Grassmannian of time--like 3--planes in 
$H^2(X,{\mathbb{R}})$ modulo $O(3,19,{\mathbb Z})$, up to a positive real
parameter which represents the volume of $X$. 
Actually, since we are interested in the case in which the $B$ field
vanishes~\footnote{See \cite{aspin1,aspin} for the subtle distinction between
the role of the $B$ field in orbifold conformal field theories and in enhanced
symmetry theories.}, we can simply identify the two moduli spaces. 

The phenomenon of enhancement of symmetry is based on the existence of the
roots of length --2, which, according to the previous subsection, correspond
to massless non--abelian gauge fields.
These correspond to homology 2--spheres $C$ in $X$ of self--intersection
$C\cdot C=-2$. 
Moreover, since they have components only in the space--like part of the 
lattice, they are orthogonal to the time--like 3--plane spanned by the 
holomorphic two--form $\Omega$ and by the K\"ahler form $\omega$ of $X$.
Now $\omega\cdot C$
measures the area of $C$, and, due to orthogonality, $\omega\cdot C=0$. 
Therefore the roots of length --2 correspond to spheres of shrinking area.
That is, our K3 will contain orbifold points \cite{witten}. This suggests a
physical picture of the origin of the enhanced symmetry: the zero mass states
are generated by 2--branes of type IIA, wrapped around the shrinking cycles.

The same can be done for the heterotic string compactified on $T^2$,
which is expected to be dual to F--theory compactified on an elliptically 
fibered K3,
\cite{vafa},\cite{vm}. With respect to the previous case, we have now a 
smaller moduli space ${\cal M}_{{\bf h},2}$ on the heterotic side.
Looking at the F--theory side, we can be more concrete. 
In fact ${\cal M}_{{\bf h},2}$ is isomorphic to the moduli space 
$\bar {\cal M}_U$ of algebraic K3's whose Picard lattice contains the
hyperbolic plane $U$. It can be shown (see e.g. \cite{aspin}, p. 77) that
this condition on $Pic(X)$ implies that $X$ is an elliptic fibration with 
a section.
Of course, in this case too we can repeat what we said above for the 
IIA theory compactified on K3. We expect the symmetry enhancement to
occur in correspondence with collapsing 2--cycles, i.e. when two or 
more singular fibres of type $I_1$ collide.
The locations of a singular fibre on the base represents the position 
of a D--7--brane of IIB theory (F--theory is by definition a 
realization of such a non--trivial configuration). This fact lends itself 
to a string theory interpretation of the enhancement of symmetry:
the massless vector states correspond to the string modes that become
massless when two or more D--7--branes collide.
However, in the F--theory case, we are in the condition to say much more: 
we can show that {\it the geometry of singular K3's contains 
enough information to allow us to reconstruct, in correspondence with 
the singular points, the expected non--abelian data, i.e. a copy of the 
Lie algebra associated to the given Dynkin diagram}. This will be the 
subject of the next section. 
As we will see, this construction actually extends also to the case of the 
IIA theory.

\section{From elliptic K3 geometry to enhanced gauge symmetry}

To start with, let us be more specific about the moduli space $\bar {\cal M}_U$
of elliptic K3's with a section. These are the K3's in which the Picard 
lattice is constrained to contain
the hyperbolic lattice $U$ spanned by a fibre $F$ and the section $\Sigma$.
Notice that every such K3 admits a Weierstrass presentation.
It is known that the locus of the the varieties $X$ with $Pic(X)=U$ is an open
smooth subvariety ${\cal M}_U\subset \bar{\cal M}_U$.
The generators of $H^2(X,{\mathbb Z})$ are the first Chern classes of 22
smooth line bundles over $X$. Among these two, namely those living in 
$H^{2,0}(X,{\mathbb C})\oplus H^{0,2}(X,{\mathbb C})
\subset H^2(X,{\mathbb Z})\otimes{\mathbb C}$, are Chern
classes of line bundles which are never algebraic, while the two line 
bundles corresponding to $F$ and $\Sigma$ are always algebraic. Notice 
also that $F$ cannot be contracted because it is not exceptional (i.e. $F\cdot 
F=0$), while contracting $\Sigma$ one loses the elliptic fibration and 
then leaves the moduli space $\bar{\cal M}_U$. The complement of these four 
generators gives us 18 line bundles 
which at generic moduli are only smooth and therefore belong to the 
transcendental lattice. These classes are our candidates to account 
for the (right--handed) abelian $U(1)^{18}$ gauge symmetry which is 
susceptible of non--abelian enhancement.

When we move to $\bar{\cal M}_U-{\cal M}_U$, the Weierstrass 
presentation gives a singular $X$: some of the transcendental cycles 
vanish but become components of the exceptional divisor $E$ on the 
resolution $\pi:\tilde X\to X$ of $X$.
As we saw in the previous section physics tells us that the enhancement of 
symmetry occurs on $X$ (not on $\tilde X$), and that the line bundles 
corresponding to the cycles which vanish on $X$ should be actually 
identified with the Cartan generators of the larger symmetry group.
Our basic observation \footnote{See \cite{Hart} for 
definitions and notations concerning sheaf theory.}
is that on $\tilde X$ each component $E_\alpha$, obtained
from the blow--up of $X$, is a divisor and, since 
$-E_\alpha\cdot E_\alpha=2$, the line bundle ${\cal O}_{\tilde 
X}(-E_\alpha)$ restricts to ${\cal O}(2)$ on $E_\alpha$.
This is good news because ${\cal O}(2)$ is the tangent sheaf $T_{E_\alpha}$ of
$E_\alpha$ and the space of its holomorphic sections
${\mathbb C}\{z^2\partial_z,\ z\partial_z,\ \partial_z\}$ is isomorphic to 
the Lie algebra $sl(2,{\mathbb C})$.
The section $z\partial_z$, which corresponds to the standard Cartan 
generator, vanishes at the nodes where $E_\alpha$ meets the other 
components of 
the singular fibre, and can be extended on $\tilde X$ as a section of 
${\cal O}_{\tilde X}(F-E_\alpha)$.
The next observation is that the direct image 
$\pi_\ast{\cal O}_{\tilde X}(F-E_\alpha)$ is not locally free on $X$: it 
is a line bundle on $X-p$, $p$ being the singular point, while its stalk
at $p$ is generated as an ${\cal O}_p$ module by the standard generators
of $sl(2,{\mathbb C})$.
This is the end of the story when we blow down only one $E_\alpha$.

To understand the general phenomenon of enhancement of symmetry 
we need the explicit realization
of the exceptional divisor $E$ in the resolution of a singularity as a
``Dynkin curve'' in the complete flag variety $\mathbb F$ associated to the
A-D-E group. We will freely use below some aspects of this construction 
and refer to \cite{S} for an expository account.
The starting point is the fact that the intersection matrix of the 
components of the exceptional curve $E$ is indeed the opposite of the 
Cartan matrix associated to the singularity.
Let us call ${\mathfrak g}$ the simple Lie algebra with such a Cartan 
matrix.
Each component $E_\alpha$ of $E$ is actually a Riemann sphere which is 
expected to correspond to the subalgebra $sl_\alpha(2)\subset{\mathfrak g}$ 
generated by a triplet associated to a root $\alpha$.

The second step comes from the embedding of $E$ as a Dynkin curve.
This goes as follows: let $r$ be the rank of $\mathfrak g$, $x\in \mathfrak 
g$ be a subregular (i.e. with a commutant $Z(x)$ of rank $r+2$) nilpotent
element, and $x,h,y$ an $sl(2)$ triplet associated to $x$. The Dynkin curve 
$E$ is the set of flags $f\in{\mathbb F}$ stabilized by
$\exp(tx),\ \forall t\in{\mathbb C}$.
As well known
\footnote{Indeed, $\mathbb F$ 
is a homogeneous $G$-space so any element $w\in{\mathfrak g}$ gives rise 
to a non-trivial fundamental holomorphic vector field on $\mathbb F$. Since 
the flag variety is compact and the fundamental vector fields span the 
tangent space to $\mathbb F$ at every point, it follows that each 
holomorphic vector field is actually fundamental.}
$\mathfrak g$ is isomorphic to the Lie algebra of the
holomorphic vector fields on $\mathbb F$.
A simple idea would be to restrict these vector fields to the
Dynkin curve $E$. However, this restriction has a non-trivial kernel: by 
definition, at least the fundamental vector field associated to $x$
restricts to zero on $E$.
By considering the infinitesimal action of an element $w\in{\mathfrak g}$ 
we see that the corresponding vector field is tangent to $E$ if and only 
if $[x,w]=\lambda x$ for some $\lambda\in{\mathbb C}$, therefore
$w+(\lambda/2)h$ commutes with $x$ showing that the space of fundamental 
vector fields tangent to $E$ is isomorphic to $Z(x)\oplus{\mathbb C}\{h\}$.
Notice moreover that $h$ does not vanish on $E$.

Our proposal to restore the entire algebra is to restrict  the
fundamental vector fields on $\mathbb F$ to a family of Dynkin curves. Since 
all Dynkin curves in $\mathbb F$ are conjugate, we look for a
subvariety ${\cal E}\to \Delta$ fibered in Dynkin curves $E_t,\; t\in 
\Delta$ (with $\Delta$ within the adjoint orbit through $x$) which is 
minimal with respect to the properties that:

\noindent{\bf P1:}{ \it no fundamental vector field vanishes 
identically on $\cal E$,}

\noindent{\bf P2:}{ \it the space of holomorphic sections of a 
subsheaf of $i^\ast T{\mathbb F}$ is isomorphic to $\mathfrak g$,}

\noindent where we denote by $i:{\cal E}\to {\mathbb F}$
the embedding of the family.

An explicit construction of $\cal E$ runs as follows:

\medskip

\noindent{\bf Proposition.} {\it Let $x,h,y$ be an $sl(2)$ triplet
associated to $x$ and let $E$ be the Dynkin curve stabilized by $x$.
The infinitesimal family ${\cal E}=\cup_{t\in{\mathbb C}}\exp(ty)\cdot E\ 
(mod\; t^2)$ satisfies the property {\bf P1}.}

\medskip

\noindent{\bf Proof.} By the above description, a fundamental vector 
field vanishing on $E$ vanishes on $\cal E$ as well if and only if it 
commutes with the entire triplet, and hence it belongs to the "reductive 
centralizer" ${\mathfrak c}=Z(x)\cap Z(y)$.
It is known \cite{S} that $\mathfrak c$ is zero for the type D and E
algebras, while it is one dimensional for the singularities of type A.
In the latter case, taking $x$ to be the standard subregular nilpotent
element (see p.87 of \cite{S}), a generator for $\mathfrak c$ reads
$c=diag(r,-1,...,-1)$ and the corresponding vector field does not vanish
on $E$. Indeed, the explicit realization of the Dynkin curve given in 
p.88 of \cite{S} shows that the vector field associated to $c$ is nonzero
at least on the component $E_\alpha$ with $\alpha=L_1-L_2$ being the
``first'' simple root (see \cite{FH} for notations). \hfill$\square$ 

\medskip

Let $\tilde{\cal F}$ be the sheaf of sections of $(i^*T{\mathbb 
F})(-E)$ on $\cal E$. A direct computation 
shows that $H^0({\cal E},\tilde{\cal F})= {\mathfrak g}$:

\medskip

\noindent{\bf Proposition.} {\it The family $\cal E$ satisfies {\bf P2}.}

\medskip

\noindent{\bf Proof.} The space $H^0({\cal E},i^\ast T{\mathbb F})$
of holomorphic sections of the restriction of the tangent bundle to $\cal E$
is generated over ${\mathbb C}[t]/t^2$ by the fundamental vector fields, 
hence it is isomorphic to $t{\mathfrak g}\oplus {\mathfrak z}$, ${\mathfrak z}=
{\mathfrak g}/\ker(i^\ast:{\mathfrak g}\to H^0(E,i^\ast T{\mathbb F}))$
being the space of fundamental vector fields not vanishing on $E$.
\hfill $\square$

\medskip

We can get rid of $\cal E$ by projecting $\varpi:{\cal E}\to E$ on $E$ and 
taking the direct image sheaf ${\cal F}=\varpi_\ast\tilde{\cal F}$.
To make contact with enhancement of symmetry we simply embed the Dynkin 
curve $E$ as the exceptional divisor of $\pi:\tilde X\to X$, consider the 
direct image $\tilde{\cal G}$ of $\cal F$ under the embedding, blow down 
$\tilde X$ to $X$ and take the direct image ${\cal G}=\pi_\ast\tilde{\cal 
G}$. This is a skyscraper sheaf with stalk the Lie algebra $\mathfrak g$ 
supported at the singular point $p\in X$.

\section{Final comments}

The results of the previous section refer to elliptically fibered K3's with
a section that is, specifically, to the F--theory case.
However the construction is local around the singularity and can be applied 
to a generic complex surface, in particular to the compactification of
IIA theory on singular K3's.
A generic K3 is not elliptically fibered and therefore the line bundle
${\cal O}(F-E_\alpha)$ does not exist. However, the construction really 
depends only on the fact that the exceptional divisor associated to a
singularity of type A-D-E is a Dynkin curve. Of course this does not 
rely on the presence of an elliptic fibration and continues to be true even
if the surface is not algebraic.
In both cases the construction above produces a skyscraper sheaf of Lie 
algebras on singular K3's.
From the point of view of space-time we have projections $q_1:X\times 
{\mathbb R}^6\to X$ and $q_2:X\times {\mathbb R}^n\to{\mathbb R}^n$ 
($n=6,8$) and $q_{2\ast}q_1^\ast{\cal G}$ is now a trivial sheaf of Lie
algebras on the noncompact part of space-time.
In conclusion, we can reconstruct out of the singular K3 a bundle of Lie 
algebras on ${\mathbb R}^n$ which is an essential ingredient to start the
study of the duality with the heterotic string.

Of course one would like to retrieve, in correspondence with symmetry 
enhancement, the full non--abelian framework of a gauge theory, 
including the gauge bosons with values in $\mathfrak g$. This however requires
some additional information, which is not encoded in the geometry of the
compactification space: in particular we need the notion of one--form in the
uncompactified space. This goes beyond the scope of this paper, so we limit 
ourselves to a few words in the case of IIA theory. Here the additional
ingredient we need comes from physics: it is the IIA theory 3-form which, on 
$X\times{\mathbb R}^6$, has a Kunneth component of degree $(2,1)$. This 
component can be written as a superposition of harmonic forms on $X$ 
whose coefficients are 1-forms on ${\mathbb R}^6$. When $X$ is singular 
one can resolve the singularities, work on the smooth model $\tilde X$ 
and take 3-forms with coefficients in $\tilde{\cal F}$. This may be a 
suggestion to get the algebra-valued 1-forms on ${\mathbb R}^6$.

\section*{Acknowledgements}

We would like to thank U. Bruzzo, I. Dolgachev, B. Dubrovin, C. Gomez,
D. Hernandez Ruiperez, J.M. Mu\~noz Porras for many useful discussions.
The work of C. Reina and A. Zampa has been partially supported by the
italian MURST and by the GNFM/CNR.
L. Bonora has been partially supported by the EC TMR Programme, grant 
FMRX-CT96-0012, and by the Italian MURST for the
program ``Fisica Teorica delle Interazioni Fondamentali''.

\section*{Appendix}

In this Appendix we consider the explicit example of compactification of the 
heterotic string on the torus $T^2$ with background $g_{11}=g_{22}= 1/4,
g_{12}=0, B_{12}=0$ (see subsection 2.1.) and construct an elliptic K3 with
the same symmetry enhancement.  
Let us introduce the notation:  ${\bf n} =(n^1,n^2)$,
${\bf m} = (m_1,m_2)$ and ${\bf A}^I = (A^I_1, A^I_2)$. Then the conditions for
the bosonic massless states are 
\a 
&&0 = - \frac {3}{2} + N_{NS} + \tilde N_A +
\frac {1}{4} {\bf n} \cdot {\bf n} + ( {\bf m}+{\bf A}^I (m^I - {1\over 2}{\bf
A}^I\cdot {\bf n}))^2 + {1\over 2} (m^I - {\bf A}^I\cdot {\bf n})^2\0\\
&&{1\over 2}= - N_{NS} + \tilde N_A + {\bf n} \cdot ( {\bf m}+{\bf A}^I (m^I -
{1\over 2}{\bf A}^I\cdot {\bf n})) + {1\over 2} (m^I - {\bf A}^I\cdot {\bf
n})^2,\label{NSA} 
\b 
where 
\a 
&&N_{NS} = \sum_{n=1}^{\infty} (\alpha_{-n}^\lambda
\alpha_n^\lambda+ \alpha_{-n}^i \alpha_n^i) + \sum_{r= {1\over 2}}^\infty (r
b_{-r}^\lambda b_r^\lambda + r b^i_{-r}b^i_r)\0\\ &&N_{A} = \sum_{n=1}^{\infty}
(\alpha_{-n}^\lambda \alpha_n^\lambda+ \alpha^i_{-n}\alpha^i_n) + \sum_{n=
1}^\infty (n d_{-n}^\lambda d_n^\lambda + nd^i_{-n}d^i_n)\0\\ &&\tilde N_{A} =
\sum_{n=1}^{\infty} (\tilde \alpha_{-n}^\lambda \tilde \alpha_n^\lambda+ \tilde
\alpha_{-n}^i \tilde \alpha_n^i) + \sum_{n= 1}^\infty \alpha^I_{-n} 
\alpha_n^I. \label{NNSA} 
\b 
Here $\lambda$ denotes the uncompactified dimensions and $i=1,2$ the
compact ones.  These indices, when repeated, are supposed to be summed
over.

Now let us restrict the values of the gauge fields to $A_i^I= \delta_i^I$. The
massless vector states are 18 right states which come from the conditions 
\a
N_{NS}= {1\over 2}, \quad \tilde N_A =1, \quad L^i g_{ij} L^j =0, \quad \tilde
L^i g_{ij}\tilde L^j +L^IL^I =0\0 .
\b 
These are the states (\ref{cartan}).  More
right massless vector states are given by tensoring $b_{-{1\over
2}}^\lambda|0>_R$ with the scalars corresponding to points of length --2 
in the left--handed lattice, i.e.  states obtained by imposing the conditions
\a 
N_{NS}=
{1\over 2}, \quad \tilde N_A =0, \quad L^i g_{ij} L^j =0, \quad \tilde L^i
g_{ij}\tilde L^j +L^IL^I =2.\label{length} 
\b 
There are altogether 612 such states, which
together with the 18 states (\ref{cartan}), form the adjoint representation of
SO(36).  The 18 states are the Cartan subalgebra generators.
Notice that the massless vector states not belonging to the Cartan 
subalgebra come from points of length --2 in the lattice (\ref{lattice}),
due to (\ref{quadratic}) and (\ref{length}).

There are also left massless vector states.  Two of them come from the
conditions 
\a N_{NS}= {1\over 2}, \quad \tilde N_A =1, \quad L^i g_{ij} L^j =0,
\quad \tilde L^i g_{ij}\tilde L^j +L^IL^I =0\0 ,
\b 
i.e.  they correspond to the states (\ref{cartan'}).  There are no more 
massless vector states.  Therefore at the point of the moduli space 
determined by $g_{11}=g_{22}= 1/4, g_{12}=0$ and $A_i^I= \delta_i^I$, the 
symmetry of the heterotic theory compactified on $T^2$ is enhanced to 
SO(36)$\times$U(1)$^2$.

Let us see the same enhancement of symmetry on the F--theory side.
An elliptically fibered K3 surface with a singularity of type $D_{18}$ is, 
for example, the one explicitely given by the following Weierstrass 
presentation: $y^2=4x^3-g_2x-g_3$, where
$$g_2=4^{1/3}(18z_0^8+30z_0^6z_1^2+12z_0^4z_1^4+3z_0^2z_1^6),$$
and
$$g_3=-(63z_0^{11}z_1+70z_0^9z_1^3+42z_0^7z_1^5+12z_0^5z_1^7+2z_0^3z_1^9).$$
We have
$$\delta=g_2^3-27g_3^2=23328z_0^{24}+9477z_0^{22}z_1^2+2916z_0^{20}z_1^4,$$
showing that this surface has a singularity of type $D_{18}$ over
$[z_0:z_1]=[0:1]\in{\mathbb P}^1$ (and four fibres of type $I_1$ in the
Kodaira classification).


\begin{thebibliography}{}

\bibitem{witten1} E.Witten, {\it Toroidal compactification without vector
structure}, hepth/9712028
\bibitem{narain}
K.S.Narain, {\it New heterotic string theories in uncompactified dimensions
$<10$} Phys.Lett. 169B (1986) 41.\\
K.S.Narain, M.H.Sarmadi and E.Witten, {\it A note on toroidal compactification 
of heterotic string theory}, Nucl.Phys.B279 (1987) 369.
\bibitem{ginsparg} P.Ginsparg, {\it Comments on toroidal compactifications
of heterotic superstrings}, Phys.Rev. D35 (1987) 648.
\bibitem{GPR} A.Giveon, M.Porrati and E.Rabinovici, {\it Target space duality 
in string theory}, Phys.Rep.244 (1994) 77,  hepth/9401139.
\bibitem{witten} E.Witten, {\it String theory dynamics in various dimensions}
Nucl.Phys.B433 (1995) 85, hepth/9503124.
\bibitem{aspin} P.S.Aspinwall, {\it K3 surfaces and string dualities}, 
hepth/9611137.
\bibitem{aspin1} P.S.Aspinwall, {\it Enhanced gauge symmetries and K3 
surfaces}, Phys.Lett.B357 (1995) 329-334, hepth/9507012.
\bibitem{vafa} C.Vafa, {\it Evidence for F--theory}, Nucl.Phys. B469 (1996) 
403.
\bibitem{vm} C.Vafa and D.R.Morrison, {\it Compactifications of F--theory on 
Calabi-Yau threefolds} --I and --II, hepth/9602114 and hepth/9603161. 
\bibitem{S} P.Slodowy, {\it Simple singularities and simple algebraic groups},
Lectures Notes in Mathematics vol. 815, Springer Verlag, Berlin 1980. 
\bibitem{FH} W.Fulton and J.Harris, {\it Representation theory. A first
course} , Graduate Texts in Mathematics vol. 129, Springer Verlag,
New York 1991.
\bibitem{Hart} R.Hartshorne, {\it Algebraic Geometry}, Graduate Texts in
Mathematics vol. 52, Springer Verlag, New York 1977
\end{thebibliography}
\end{document}